# Exact Solutions of the Klein-Gordon Equation for the Rosen-Morse type Potentials via Nikiforov-Uvarov Method

A. Rezaei Akbarieh and H. Motavali[1]
1 Faculty of Physics, University of Tabriz, Tabriz 51664, Iran

The exact solutions of the one-dimensional Klein-Gordon equation for the Rosen-Morse type potential with equal scalar and vector potentials are presented. First we briefly review Nikiforov-Uvarov mathematical method. Using this method, wave-functions and corresponding exact energy equation are obtained for the s-wave bound state. It has been shown that the results for Rosen-Morse type potentials reduce to standard Rosen-Morse well and Eckart potentials in the special case. The PT-symmetry for these potentials is also considered.

## 1. Introduction

In recent years, there has been an increased interest in finding exact solutions of the Klein-Gordon (KG) equation [1-5]. It is well known that these solutions play an essential role in the relativistic quantum mechanics for some physical potentials of interest [6-10]. Some known potentials have been considered in the literature [1-5]. Generalized Hulthen potential which is reducible to the standard Hulthen potential, Woods-Saxon potential, exponential type screened potential are studied in Ref. [11]. Relativistic bound states of the standard Hulthen potential has also been presented by Dominguez-Adam [12] and Chetouani et al. [13]. Moreover, Rao and Kagali [14] found out the relativistic solutions of the one dimensional KG equation for the exponential potential, and Znojil [15] obtained the solutions of Schrödinger equation in the non-relativistic region. However, there is no explicit expression for the energy eigenvalue in both relativistic and non-relativistic regions.

In this paper first we briefly review Nikiforov-Uvarov (NU) mathematical method which is based on solving the second-order linear differential equation in which the differential equation is reduced to a generalized equation of hyper-geometric type [16]. Next, this method is used to solve the KG equation for Rosen-Morse type potential with equal scalar and vector potentials. The exact solution for the s-wave bound state energy eigenfunction and corresponding energy equation are obtained. Other Rosen-Morse type potentials are also discussed. The PT-symmetry for these potentials is also considered and some figures for a specific kind of solutions are presented.

## 2. NU Method

NU method is based on reducing the second order linear differential equation to a generalized equation of hyper-geometric type [16]. This method provides exact solutions in terms of special orthogonal functions as well as corresponding eigenvalues. In both relativistic and non relativistic quantum mechanics, the equation with a given real or complex potential can be solved by this method. Here, we use this method for solving KG equation with equal scalar and vector potentials. By

---

1- E-mail address: Motavalli@tabrizu.ac.ir
 Tel: +98-4113392664
 Fax: +98-4113341244

introducing an appropriate coordinate transformation $s = s(x)$, one can rewrite this equation in the following form

$$\psi''(s) + \frac{\tilde{\tau}(s)}{\sigma(s)}\psi'(s) + \frac{\tilde{\sigma}(s)}{\sigma^2(s)}\psi(s) = 0 \tag{1}$$

where $\sigma(s)$ and $\tilde{\sigma}(s)$ are polynomials of degree two at most, and $\tilde{\tau}(s)$ is a polynomial of degree one at most [16]. Now, if one takes the following factorization

$$\psi(s) = \phi(s)y_n(s)$$

Eq. (1) reduces to a hyper-geometric type equation of the form

$$\sigma(s)y_n''(s) + \tau(s)y_n'(s) + \lambda y_n(s) = 0$$

where $\tau(s) = \tilde{\tau}(s) + 2\pi(s)$ satisfy the condition $\tau'(s) < 0$, and $\pi(s)$ is defined as

$$\pi(s) = (\frac{\sigma'(s) - \tilde{\tau}(s)}{2}) \pm \sqrt{(\frac{\sigma'(s) - \tilde{\tau}(s)}{2})^2 - \tilde{\sigma}(s) + \kappa\sigma(s)} \tag{2}$$

here $\kappa$ is a parameter. Determining of $\kappa$ is the essential point in the calculation of $\pi(s)$. It is simply defined by setting the discriminant of the square root to zero [16]. Therefore, one gets a general quadratic equation for $\kappa$. The values of $\kappa$ can be used for calculation of energy eigenvalue by using

$$\lambda = \kappa + \pi'(s) = -n\tau'(s) - \frac{n(n-1)}{2}\sigma''(s) \qquad n = 0,1,2,.... \tag{3}$$

Polynomial solutions $y_n(s)$ are given by Rodrigues relation

$$y_n(s) = \frac{B_n}{\rho(s)}(\frac{d}{ds})^n[\sigma^n(s)\rho(s)], \qquad n = 0,1,2,... \tag{4}$$

in which $B_n$ is a normalization constant and $\rho(s)$ is the weight function satisfying

$$(\sigma\rho)' = \tau\rho. \tag{5}$$

On the other hand, the function $\phi(s)$ satisfies the condition

$$\phi'(s)/\phi(s) = \pi(s)/\sigma(s). \tag{6}$$

## 3. Exact Bound-State Solutions

Let us begin with time independent KG equation for a spinless particle of rest mass $m$ with the natural units $\hbar = c = 1$

$$\psi''(x) + [(E - V(x))^2 - (m + S(x))^2]\psi(x) = 0 \tag{7}$$

٢

where $E$, $V(x)$ and $S(x)$ are the relativistic energy of the particle, vector and scalar potentials, respectively. Recently, interest in the solutions of this equation with equal scalar and vector potentials has been increased. If we put $V(x) = S(x)$ the last equation takes the form of Schrödinger equation

$$\left[-\frac{d^2}{dx^2} + V_{eff}(x)\right]\psi(x) = \overline{E}^2\psi(x)$$

where $V_{eff}(x) = -2(E+M)V(x)$ and $\overline{E}^2 = E^2 - M^2$ are the effective potential and energy parameter respectively. Now we consider the one dimensional Rosen-Morse type potential as [17]

$$V(x) = -V_1 \sec h_q^2(\alpha x) - V_2 \tanh_q(\alpha x)$$

Where $1 \geq q \geq -1$. Here we use the following deformed hyperbolic functions [18]

$$\sinh_q x = \frac{e^x - qe^{-x}}{2}, \quad \cosh_q x = \frac{e^x + qe^{-x}}{2}, \quad \tanh_q x = \frac{\sinh_q x}{\cosh_q x}$$

$$\coth_q x = \frac{\cosh_q x}{\sinh_q x}, \quad \sec h_q x = \frac{1}{\cosh_q x}, \quad \cos ech_q x = \frac{1}{\sinh_q x}.$$

The deformation parameter is taken positive and real in Ref. [19] and the energy equation and the corresponding wave function are obtained. Extension to complex parameter has been considered in Refs. [20-21].

The deformed hyperbolic functions allow us to investigate the effect of the deformation parameter $q$ on the energy levels and the corresponding wave functions. Moreover, this potential is reduced to some standard potential by choosing an appropriate value for the deformation parameter. The deformed Rosen-Morse potential is transformed into the standard Rosen-Morse potential form for $q=1$, into the standard Eckart potential for $q=-1$ and into the exponential potential for $q=0$ apart from a constant shift. By using a coordinate translation transformation, it has been shown that deformed hyperbolic potentials can be transformed into the corresponding non-deformed ones [23].

Fig. 1 shows the variation of this potential in terms of $x$ for different values of $q$ with $V_2 = -\frac{1}{3}V_1$ and $\alpha = V_1 = 1$.

Using this potential Eq. (7) takes the form

$$\psi''(x) + \left[\overline{E}^2 + \overline{V}_1 \sec h_q^2(\alpha x) + \overline{V}_2 \tanh_q(\alpha x)\right]\psi(x) = 0$$

where $\overline{V}_1 \equiv 2(E+M)V_1$ and $\overline{V}_2 \equiv 2(E+M)V_2$. It is straightforward to show that by introducing a new variable $s = \tanh_q(\alpha x)$, the last equation takes the form

$$\psi''(s) - \frac{2s}{1-s^2}\psi'(s) - \frac{1}{(1-s^2)^2}\left[-\varepsilon^2 s^2 + \gamma^2 s + \varepsilon^2 - \beta^2\right]\psi(s) = 0$$

where, we have used

$$\varepsilon^2 = -\frac{\overline{V_1}}{q^3\alpha^2}, \quad \gamma^2 = -\frac{\overline{V_2}}{q^2\alpha^2}, \quad \beta^2 = -\frac{\overline{E}^2}{q^2\alpha^2}.$$

Now, by comparing the last equation and Eq. (1), we get

$$\tilde{\tau}(s) = -2s, \quad \sigma(s) = 1 - s^2, \quad \tilde{\sigma}(s) = -\varepsilon^2 s^2 + \gamma^2 s + \varepsilon^2 - \beta^2.$$

Substituting them into relation (2) leads to

$$\pi(s) = \pm\sqrt{(\varepsilon^2 - \kappa)s^2 - \gamma^2 s + (\beta^2 - \varepsilon^2 + \kappa)}$$

then, one gets two possible functions for each root $\kappa$ as

$$\pi(s) = \pm(\mu s - \nu) \quad \text{for} \quad \kappa_\pm = (2\varepsilon^2 - \beta^2 \pm \sqrt{\beta^4 - \gamma^4})/2$$

where

$$\mu^2 = (\beta^2 - \sqrt{\beta^4 - \gamma^4})/2, \quad \nu^2 = (\beta^2 + \sqrt{\beta^4 - \gamma^4})/2.$$

Imposing $\tau'(s) < 0$, we obtain the energy eigenvalue equation from (3) for $\kappa_-$ as follows

$$\varepsilon^2 - \frac{1}{2}\beta^2 - \frac{1}{2}\sqrt{\beta^4 - \gamma^4} - \sqrt{\frac{1}{2}\beta^2 - \frac{1}{2}\sqrt{\beta^4 - \gamma^4}} = 2n(\mu+1)s + 2n\nu + n(n-1)$$

or explicitly

$$\overline{E}^2 = -\frac{\overline{V_2}^2}{\alpha^2}(2n+1-\sqrt{1+\frac{4\overline{V_1}}{q\alpha^2}})^{-2} - \frac{\alpha^2}{4}(2n+1-\sqrt{1+\frac{4\overline{V_1}}{q\alpha^2}})^2. \tag{8}$$

From (5) it can be shown that the weight function $\rho(s)$ is

$$\rho(s) = (1-s)^{\mu+\nu}(1+s)^{\mu-\nu}$$

and by substituting into the Rodrigues relation (4) one gets

$$y_n(s) = B_n(1-s)^{-(\mu+\nu)}(1+s)^{-(\mu-\nu)}(\frac{d}{ds})^n\left[(1-s^2)^n(1-s)^{-(\mu+\nu)}(1+s)^{-(\mu+\nu)}\right]$$
$$= B_n P_n^{(\mu+\nu),(\mu-\nu)}(s)$$

where, $B_n$ stands for the normalization constant.
The other part of the wave function in Eq. (6) is simply found as

$$\phi(s) = (1-s)^{\frac{1}{2}(\mu+\nu)}(1+s)^{\frac{1}{2}(\mu-\nu)}.$$

ε

Finally, by multiplying the two parts, one obtains

$$\psi(s) = (1-s)^{\frac{1}{2}(\mu+\nu)}(1+s)^{\frac{1}{2}(\mu-\nu)} P_n^{(\mu+\nu,\mu-\nu)}(s)$$

or equivalently

$$\psi(\tanh_q(\alpha x)) = (1-\tanh_q(\alpha x))^{\frac{1}{2}(\mu+\nu)}(1+\tanh_q(\alpha x))^{\frac{1}{2}(\mu-\nu)} P_n^{(\mu+\nu,\mu-\nu)}(\tanh_q(\alpha x)). \tag{9}$$

This wave-function and energy equation (8) are exactly consistent with the results of Refs. [7, 19-20].

## 4. Results and Discussion

One can choose an appropriate value for the parameters in the Rosen-Morse type potential to obtain the well-known potentials.

### 4.1 Rosen-Morse Well

For $q=1$ and $V_2 \to -V_2$ the Rosen-Morse type potential transforms to the standard Rosen-Morse well as follows

$$V(x) = -V_1 \sec h^2(\alpha x) + V_2 \tanh(\alpha x). \tag{10}$$

This potential is widely used in studying vibrations of polyatomic molecules. As an example, the vibrational states of $NH_3$ has been considered in Ref. [17]. The corresponding energy equation and wave function for (10) are

$$\overline{E}^2 = -\frac{\overline{V_2}^2}{\alpha^2}(2n+1-\sqrt{1+\frac{4\overline{V_1}}{\alpha^2}})^{-2} - \frac{\alpha^2}{4}(2n+1-\sqrt{1+\frac{4\overline{V_1}}{\alpha^2}})^2$$

and

$$\psi(\tanh \alpha x) = (1-\tanh \alpha x)^{\frac{1}{2}(\mu+\nu)}(1+\tanh \alpha x)^{\frac{1}{2}(\mu-\nu)} P_n^{(\mu+\nu,\mu-\nu)}(\tanh \alpha x)$$

respectively. We have plotted the variation of the energy in Figs. 2 and 3, for four different values of $q$ and $n$.

### 4.2 Eckart Potential

For $q=-1$ and $V_1 \to -V_1$ the Rosen-Morse type potential changes to the standard Eckart potential [23]

$$V(x) = V_1 \sec h^2(\alpha x) - V_2 \tanh(\alpha x).$$

The corresponding energy equation and wave function for this potential are

$$\overline{E}^2 = -\frac{\overline{V}_2^2}{\alpha^2}(2n+1-\sqrt{1-\frac{4\overline{V}_1}{\alpha^2}})^{-2} - \frac{\alpha^2}{4}(2n+1-\sqrt{1-\frac{4\overline{V}_1}{\alpha^2}})^2,$$

and

$$\psi(\coth\alpha x) = (1-\coth\alpha x)^{\frac{1}{2}(\mu+\nu)}(1+\coth\alpha x)^{\frac{1}{2}(\mu-\nu)} P_n^{(\mu+\nu,\mu-\nu)}(\coth\alpha x)$$

respectively.

### 4.3 PT-Symmetric Rosen-Morse Well

In general, when a Hamiltonian commutes with PT, it is called PT-symmetric Hamiltonian, where P denotes parity operator (space reflection, $P : x \to -x$) and T denotes time-reversal operator $(T : i \to -i)$. If $V(x)$ under the both transformations $x \to -x$ (or $x \to \eta - x$) and $i \to -i$ behaves as $V^*(x) = V(-x)$ the potential $V(x)$ is called PT-symmetric.

Now let us return to Rosen-Morse well which under the transformation $V_2 \to iV_2$ takes the form [23]

$$V(x) = -V_1 \sec h_q^2(\alpha x) - iV_2 \tanh_q(\alpha x)$$

where $V_1 > 0$ and $q > 0$. The corresponding replacements in the energy equation for the PT-symmetric version of the Rosen-Morse well led to

$$\overline{E}^2 = \frac{\overline{V}_2^2}{\alpha^2}(2n+1-\sqrt{1+\frac{4\overline{V}_1}{q\alpha^2}})^{-2} - \frac{\alpha^2}{4}(2n+1-\sqrt{1+\frac{4\overline{V}_1}{q\alpha^2}})^2.$$

One can easily check that the corresponding wave-function with the following parameters

$$\mu^2 = (\beta^2 - \sqrt{\beta^4+\gamma^4})/2 \quad , \quad \nu^2 = (\beta^2 + \sqrt{\beta^4+\gamma^4})/2$$

is the same as (9).

### 4.4 PT-Symmetric Version of the Eckart Potential

The following replacements
$$V_1 \to -V_1 Q, \quad q \to -Q, \quad V_2 \to iV_2, \quad Q \to e^{2i\alpha\theta}$$
give PT-symmetric version of the Eckart potential [21]
$$V(x) = -V_1 Q \cos ech_Q^2(\alpha x) - iV_2 \coth_Q(\alpha x)$$
where $V_1 > 0$, $0 < \theta < \frac{1}{4}\pi$ or $-\frac{\pi}{4} < \theta < 0$.

### 5. Conclusions

In summary, using NU method we have discussed the exact solution of the Klein-Gordon equation with equal scalar and vector potentials for s-wave bound states. We have obtained the energy eigenvalue equation and eigenfunction of the Klein-Gordon equation for the Rosen-Morse type potential. In addition, as a special case, standard Rosen-Mores well and Eckart potentials were considered. Finally, the PT-symmetry for these potentials was also discussed.

**Figure Captions**

**Fig. 1.** A schematic representation of the Rosen-Morse type potential for four different values of the shape parameter $q$ for $V_2 = -\frac{1}{3}V_1$ and $\alpha = V_1 = 1$.

**Fig. 2** The variation of the ground-state energy $(n = 0)$ in terms of the coupling constant $V = V_1$ for the Rosen-Morse type potential for different values of $q$ with $V_2 = -\frac{1}{3}V_1$ and $\alpha = 1$.

**Fig. 3** The variation of the energy in terms of the coupling constant $V = V_1$ for the Rosen-Morse type potential for different values of $n$ with $V_2 = -\frac{1}{3}V_1$ and $\alpha = 1$.

**Fig. 1**

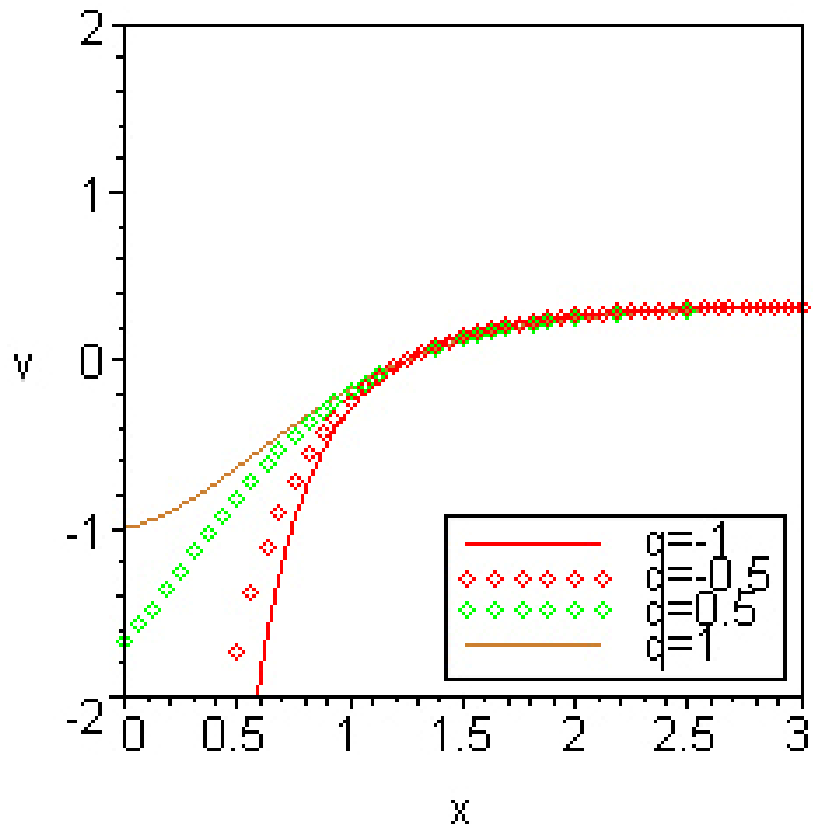

**Fig.2**

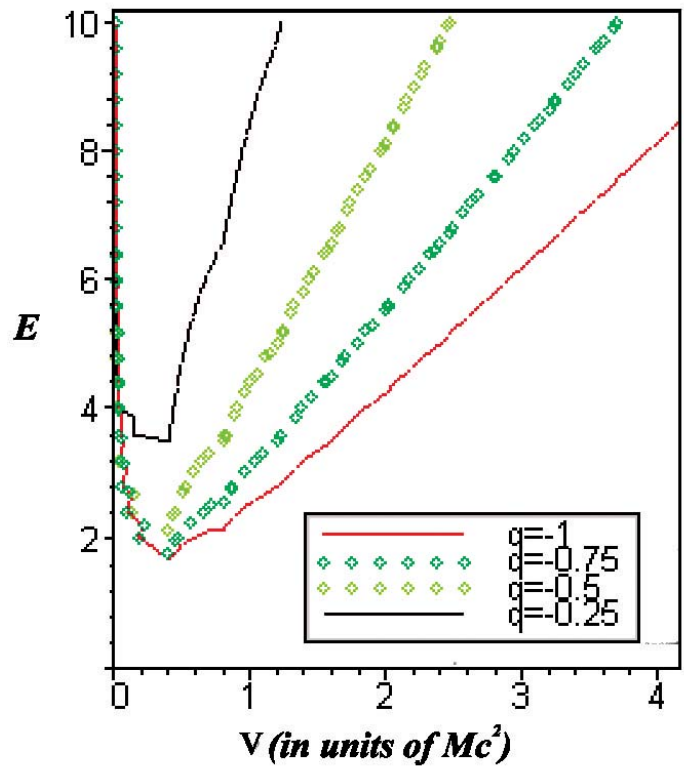

**Fig. 3**

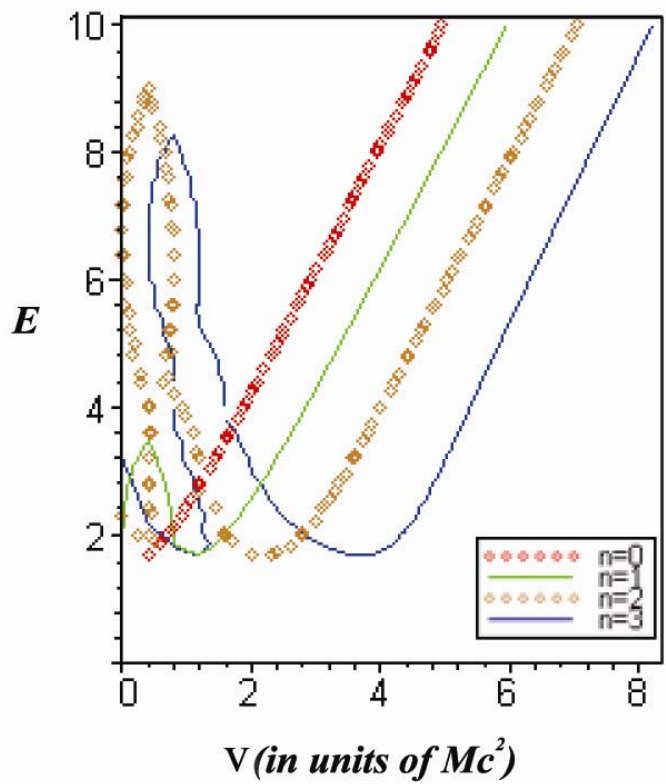